%% file: main.tex
\documentclass[10pt,conference]{IEEEtran}
\IEEEoverridecommandlockouts
\usepackage{cite}
\usepackage{amsmath,amssymb,amsfonts}
\usepackage{algorithmic}
\usepackage{graphicx}
\usepackage{textcomp}
\usepackage{xcolor}

\usepackage[ruled, linesnumbered, commentsnumbered, noend]{algorithm2e}
\usepackage{graphicx}
\usepackage{subfigure}
\usepackage{makecell}
\usepackage{multirow}
\usepackage{threeparttable}

\def\BibTeX{{\rm B\kern-.05em{\sc i\kern-.025em b}\kern-.08em
    T\kern-.1667em\lower.7ex\hbox{E}\kern-.125emX}}
\begin{document}

\newcommand{\name}{RULF}

\title{\name{}: Rust Library Fuzzing via API Dependency Graph Traversal}

\author{
\IEEEauthorblockN{Jianfeng Jiang, Hui Xu\textsuperscript{*}\thanks{\textsuperscript{*}corresponding author}, Yangfan Zhou}
\IEEEauthorblockA{School of Computer Science \\ Fudan University} 
}

\author{
\IEEEauthorblockN{
Jianfeng Jiang\IEEEauthorrefmark{1}\IEEEauthorrefmark{2},
Hui Xu\IEEEauthorrefmark{1}
Yangfan Zhou\IEEEauthorrefmark{1}\IEEEauthorrefmark{2}
}

\IEEEauthorblockA{\IEEEauthorrefmark{1}School of Computer Science, Fudan University, Shanghai, China.}

\IEEEauthorblockA{\IEEEauthorrefmark{2}Shanghai Key Laboratory of Intelligent Information Processing, Shanghai, China. \\ Email: \{jfjiang19, xuh, zyf\}@fudan.edu.cn}
}

\maketitle

\begin{abstract}
Robustness is a key concern for Rust library development because Rust promises no risks of undefined behaviors if developers use safe APIs only. Fuzzing is a practical approach for examining the robustness of programs. However, existing fuzzing tools are not directly applicable to library APIs due to the absence of fuzz targets. It mainly relies on human efforts to design fuzz targets case by case which is labor-intensive. To address this problem, this paper proposes a novel automated fuzz target generation approach for fuzzing Rust libraries via API dependency graph traversal. We identify several essential requirements for library fuzzing, including validity and effectiveness of fuzz targets, high API coverage, and efficiency. To meet these requirements, we first employ breadth-first search with pruning to find API sequences under a length threshold, then we backward search longer sequences for uncovered APIs, and finally we optimize the sequence set as a set covering problem. We implement our fuzz target generator and conduct fuzzing experiments with AFL++ on several real-world popular Rust projects. Our tool finally generates 7 to 118 fuzz targets for each library with API coverage up to 0.92. We exercise each target with a threshold of 24 hours and find 30 previously-unknown bugs from seven libraries.
\end{abstract}


\section{Introduction}

Rust is an emerging programming language that promotes memory-safety features while not sacrificing much performance. It promises developers that their programs would not suffer undefined behaviors if they do not use unsafe code. Meanwhile, the language embraces many novel features and best practices of other programming languages, such as smart pointers and RAII (Resource Acquisition Is Initialization)~\cite{jung2017rustbelt}. Due to these advantages, Rust has surged into popularity in recent years~\cite{stackoverflow2020} and has been adopted by many academic projects (\textit{e.g.,} Redleaf~\cite{narayanan2020redleaf}, Theseus~\cite{boos2020theseus}, Tock~\cite{levy2017multiprogramming}) and industrial ones (\textit{e.g.,} Intel Cloud Hypervisor VMM~\cite{intelCloudHypervisor}, TiDB~\cite{TiDB}, Occlum~\cite{occlum2020}).

Although Rust provides specially-tailored mechanisms at the language level to enhance security, many severe bugs are still reported in existing Rust projects. In particular, Advisory-DB~\cite{Advisory-db} and Trophy-Case~\cite{trophy-case} are two well-known public repositories with hundreds of bugs found in Rust projects. An interesting phenomenon of these bugs is that most of their host programs are libraries~\cite{xu2020memory}. Such bugs can only be triggered by composing a program with a specific usage of the library APIs. Since Rust emphasizes software security and robustness, it is essential to hunt these library bugs. Unfortunately, we still lack effective tools for examining the robustness of APIs. For example, fuzzing~\cite{mcnally2012fuzzing} and symbolic execution~\cite{symbolic2013} generally require executable problems; formal verification~\cite{jung2017rustbelt} cannot be fully automated for third-party library APIs.

In this paper, we aim to bridge the gap between Rust library fuzzing and existing fuzzing tools. Fuzzing is a widely-adopted testing method that exercises a program by automatically generating inputs in a random or heuristic way. However, a major problem confronted by Rust library fuzzing is the absence of fuzz targets. A fuzz target defines an array of bytes as input for executing a program composed with some library APIs~\cite{fuzztarget}. Fuzzing tools can mutate the input of fuzz targets to explore different paths. Existing fuzzing tools, such as AFL~\cite{AFL}, honggfuzz~\cite{Honggfuzz} and libFuzzer~\cite{LibFuzzer}, all require fuzz targets for library fuzzing, and writing fuzz targets mainly relies on human efforts. Fudge~\cite{babic2019fudge} is a recently proposed fuzz target generator for C/C++ programs by extracting code snippets from Google code bases. However, its effectiveness largely depends on the library usage and suffers substantial limitations. For example, it is not applicable to newly-released libraries or APIs; or it cannot generate fuzz targets for unused APIs, but bugs may relate to rarely-used features.

This paper investigates an automated fuzz target generation approach. Our approach pursues four objectives: \textit{validity} which means the program should be successfully compiled, \textit{effectiveness} which means the fuzz targets should be friendly to fuzzing tools in reaching high code coverage or bug finding, \textit{coverage} and \textit{efficiency} meaning the fuzz targets should cover as many APIs as possible and their set should be as small as possible. To ensure validity, we compose fuzz targets based on the API dependency graph of a given library. Since each fuzz target can be viewed as a sequence of API calls, we breadth-first search (BFS) API sequences under a length threshold on the graph. For each uncovered API (deep-API) due to the length limitation, we backward search their dependent API sequences. Finally, we refine our set of sequences to obtain a minimum subset that covers the same set of APIs.

We implement a fuzz target generator, \name{}\footnote{The acronym of \textbf{RU}st \textbf{L}ibrary \textbf{F}uzzing}. Given the API specification of a Rust library, it can generate a set of fuzz targets and seamlessly integrated with AFL++~\cite{fioraldi2020afl++} for fuzzing. We conduct experiments with 14 popular Rust libraries, including three from GitHub and eleven from crates.io (the official Rust crate registry). With a depth bound of three for BFS, we generate 7-118 fuzz targets for each library. Further, we fuzz each target with a budget of 24 hours and find 30 previously-unknown bugs in seven libraries.

We summarize the main contributions of this paper as follows:

\begin{itemize}
    \item Our work serves as a pilot study to automatically generate fuzz targets for Rust libraries. It extends the applicability of existing fuzzing techniques, and such extension is urgently needed by Rust considering its intolerance of undefined behaviors.
    \item Our proposed approach leverages a sophisticated traversal algorithm, which can achieve high API coverage with only a small set of shallow fuzz targets. Such an approach is proven effective and efficient. It can shed light to further investigation on code traversal.
    \item We have implemented an open-source prototype fuzz target generator for Rust libraries\footnote{RULF is publicly available on https://github.com/Artisan-Lab/RULF}. With the tool, we successfully find 30 previously-unknown bugs in seven popular Rust libraries.
\end{itemize}

The rest of our paper is organized as follows. Section \ref{section: Preliminary} presents our motivation and research goals. We define API dependency graph and introduce our proposed traversal algorithm in section \ref{section:API dependency graph}, followed by the implementation of our prototype tool in section \ref{section:Implementation}. Section \ref{section:experiment} presents our evaluation experiments. Section \ref{section:related work} reviews related work. We conclude our paper in section \ref{section:conclusion}. 

\section{Problem of Rust Library Fuzzing} \label{section: Preliminary}

\subsection{Motivating Example}

As Rust surges into popularity, many bugs of projects written in Rust have been reported and analyzed, such as those collected in Advisory-DB and Trophy-Case. A special characteristic of these bugs is that most of them are located in libraries~\cite{xu2020memory}. Note that testing libraries is very different from testing executable programs. For library testing, developers should compose their use cases of library APIs before testing them or base the testing on unit tests. Since there could be different ways or contexts of using an API, library testing is more challenging.

\begin{figure}[ht]
    \centering
    \subfigure[The bug detail.]{
        \label{fig:bug-analysis}
        \includegraphics[width=\linewidth]{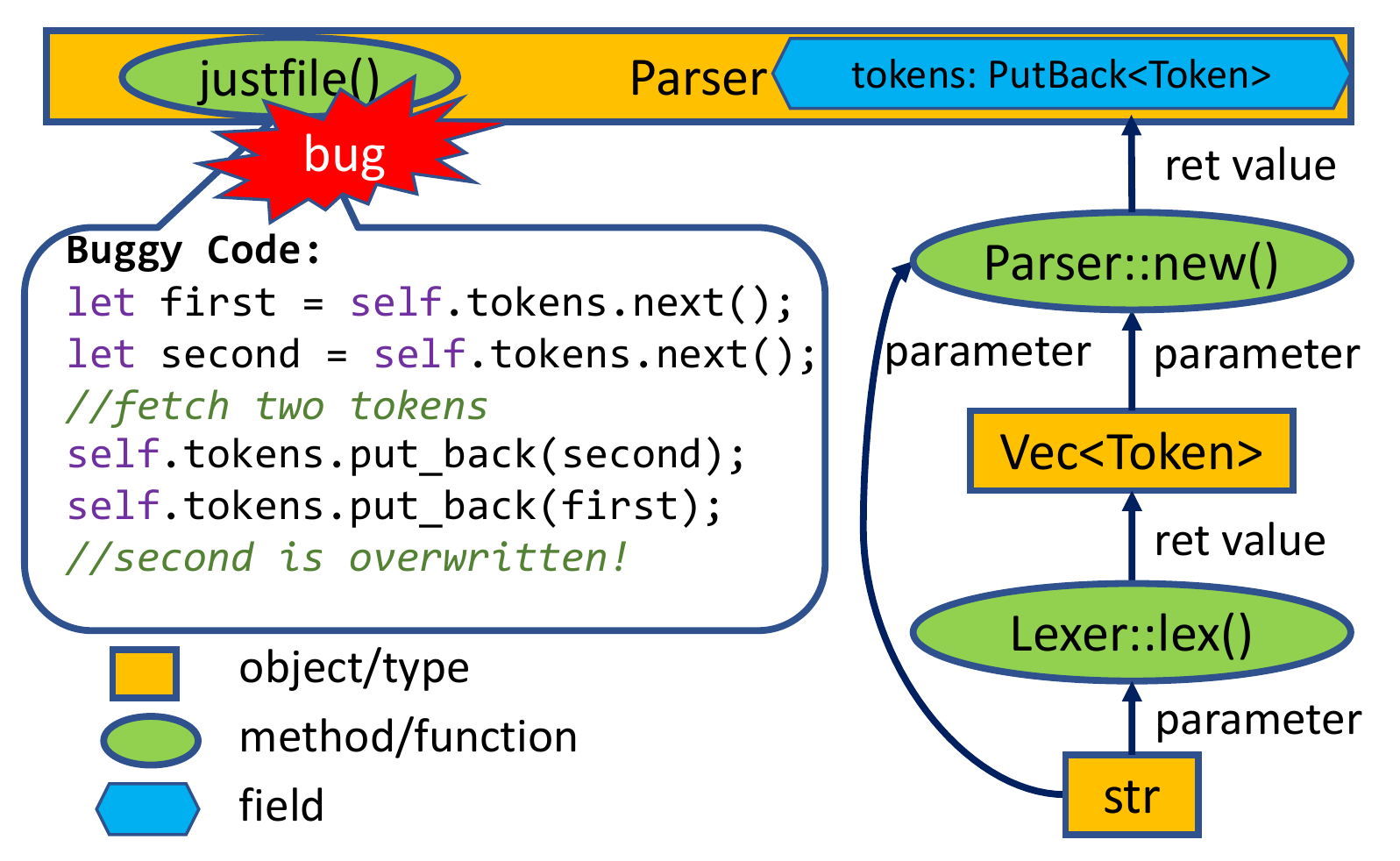}
    }
    \subfigure[Code to reproduce the bug.]{
        \label{fig:bug-reproduce}
        \includegraphics[width=\linewidth]{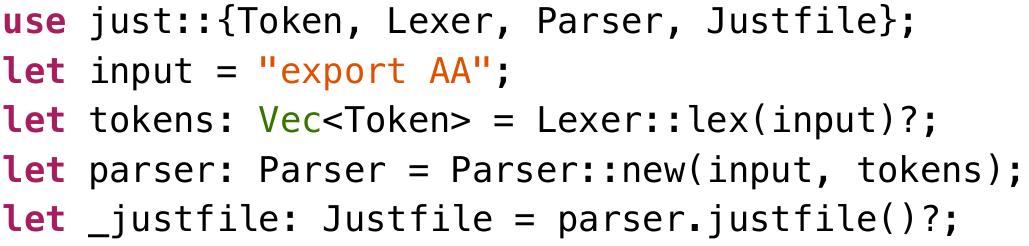}
    }
    \caption{An example bug in Rust crate \texttt{just}.}
    \label{fig:just-bug}
\end{figure}

Figure~\ref{fig:bug-analysis} demonstrates a library bug\footnote{Issue link: https://github.com/casey/just/issues/363} found in a Rust crate \texttt{just}. The term \texttt{crate} represents the minimal compilation unit in Rust. In this paper, we call a Rust library a \texttt{crate}. The bug lies in the method \texttt{justfile()}, which is defined on a struct \texttt{Parser}. The method \texttt{justfile()} uses a third-party data structure \texttt{PutBack} to store tokens. In the buggy code, the first two statements fetch two tokens, and the next two statements put the two tokens back via \texttt{put\_back()}. However, \texttt{put\_back()} can put only one single item back each time. If putting two items back, the first item will be overwritten. As a result, \texttt{justfile()} may parse an input string incorrectly and panic the application. The patch simply replaces the struct \texttt{PutBack} with another one \texttt{PutBackN} that allows putting back multiple items.

This bug had existed in the crate for over 10 versions before it was found by a fuzzing tool. Reproducing the bug requires composing a testing program for \texttt{justfile()} and some specific test cases. To this end, the testing program should create a \texttt{Parser} object via \texttt{Parser::new(\&str, Vec<Token>)}, and \texttt{Vec<Token>} can be created by calling \texttt{Lexer::lex()}. Figure~\ref{fig:bug-reproduce} presents a sample testing program based on the analysis. Specifying ``export  AA'' as the value of \texttt{input} would crash the program.

\subsection{Challenge of Fuzz Target Generation}\label{sec:challenge}

To fuzz a library program, we need a set of testing programs as fuzz targets. Such testing programs are similar to those discussed in Figure~\ref{fig:just-bug}. However, how to automatically generate such testing programs is a challenging problem. Note that there could be dozens or even hundreds of APIs for a library program, and the fuzz targets should consider each API. 

\begin{figure*}[ht]
    \centering
    \subfigure[A toy Rust crate \texttt{toylib}.]{
        \label{fig:example}
        \includegraphics[width=0.25\linewidth]{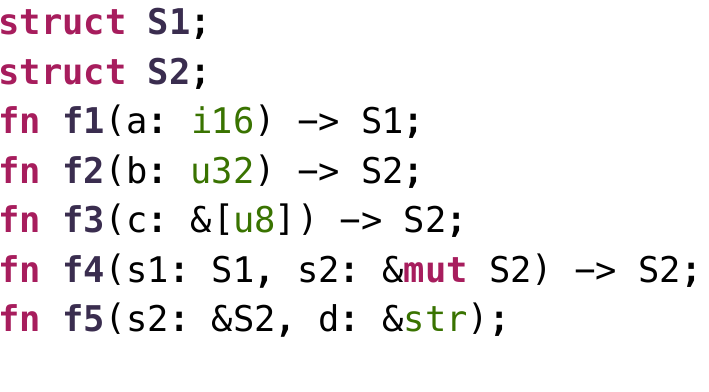}
    }
    \subfigure[Fuzz target 1.]{
        \label{fig:target1}
        \includegraphics[width=0.36\linewidth]{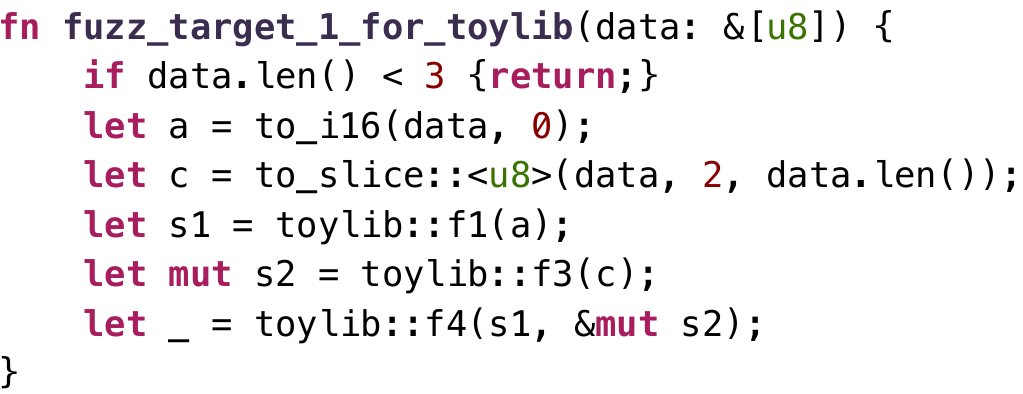}
    }
    \subfigure[Fuzz target 2.]{
        \label{fig:target2}
        \includegraphics[width=0.33\linewidth]{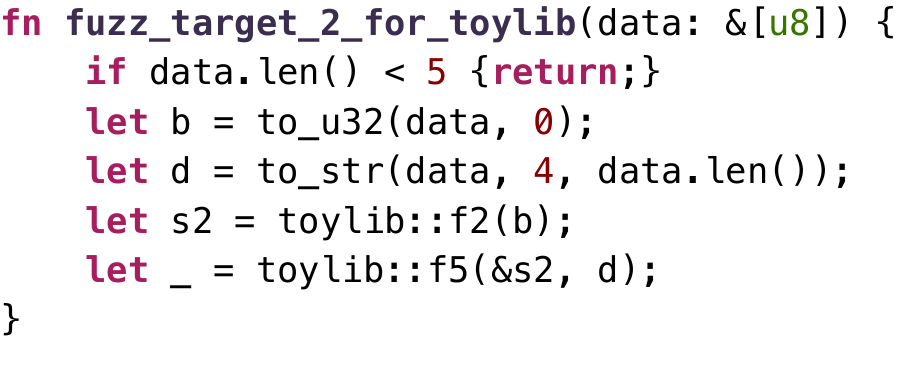}
    }
    \caption{A sample problem for fuzz target generation.}
    \label{fig:targets}
\end{figure*}

Taking the library in Figure\ref{fig:example} as an example. The crate \texttt{toylib} has two structs (\texttt{S1} and \texttt{S2}) and five public APIs (\texttt{f1} to \texttt{f5}). We may design one specific fuzz target for each API, or a large target that covers all APIs, or two potential fuzz targets (Figure~\ref{fig:target1} and Figure~\ref{fig:target2}) that can cover all APIs using each API only once. Such options are essential factors affecting the effectiveness and efficiency of library fuzzing. 

Below, we discuss four objectives for the fuzz target generation problem.

\begin{itemize}
    \item \textbf{Validity:} The synthesized fuzz targets should be able to be successfully compiled, \textit{i.e.,} all the parameters of each API should be correctly settled. \name{} achieves this objective by traversing the API dependency graph.
    \item \textbf{API coverage:} For a library program, each of APIs has a potential to contain bugs. Therefore, the set of generated fuzz targets should cover as many APIs as possible.
    \item \textbf{Efficiency:} The set of generated fuzz targets should be efficient for fuzzing. On the one hand, the number of fuzz targets should be manageable since fuzzing each target generally requires several hours. On the other hand, we should avoid employing the same precedent APIs in different fuzz targets if possible because fuzzing the same API multiple times may not be helpful.
    \item \textbf{Effectiveness:} The generated targets should be friendly to fuzzing tools for bug hunting. In order to fuzz an API effectively, the generated program should be small. Otherwise, the fuzzing effort would be wasted on the testing program or other APIs rather than the target API.
\end{itemize}

\section{Approach}\label{section:API dependency graph}

We model the problem of fuzz targets generation as an API dependency graph traversal problem. This section first defines the API dependency graph and then introduces our traversal algorithm.
 
 \subsection{API Dependency Graph}
If the return value of one API and the parameter of another API are of the same type, we can synthesize a valid program that call the second API with the return value of the first API as the parameter. In other words, we say there is one possible data flow dependency between the two APIs. Informally, an API dependency graph is a directed graph that captures all such possible data flow dependencies among the APIs of a crate.

Formally, we define an API dependency graph as a directed graph $G = (FN_m, PAR_n, PE_p, CE_q)$ over a set of APIs $\Sigma$, where:

\begin{itemize}
    \item $FN_m$ (API nodes) corresponds to all $m$ APIs in $\Sigma$;
    \item $PAR_n$ (parameter nodes) corresponds to a subset of API parameters in $\Sigma$ excluding those of primitive types. We do not consider primitive types because they can be provided by fuzzer engines.
    \item $PE_p$ (producer edges) $\subseteq$ $FN_m\times PAR_n$, where an edge $fn_i\to par_j$ implies $fn_i$ returns a value of type (or can be inferred as type) $par_j$.
    \item $CE_q$ (consumer edges) $\subseteq$ $PAR_n\times FN_m$, where an edge $par_i\to fn_j$ implies $par_i$ is a non-primitive parameter for $fn_j$. The edge is weighted if $fn_j$ requires multiple $par_i$s as its parameters.   
\end{itemize}

\begin{figure}[ht]
    \centering
    \includegraphics[width=0.8\linewidth]{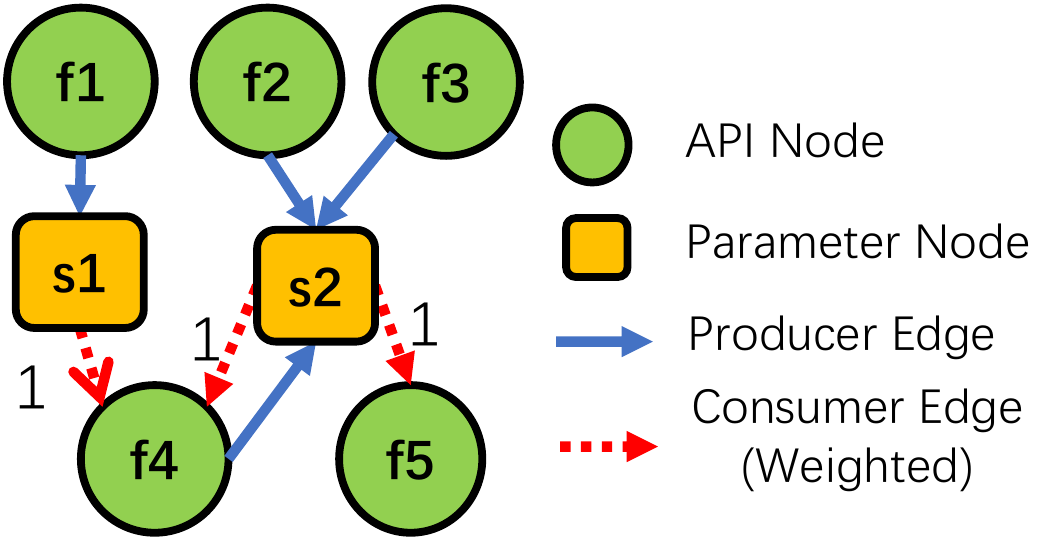}
    \caption{An API dependency graph of crate \texttt{toylib}.}
    \label{fig:ADGraph}
\end{figure}

We introduce parameter nodes into the graph to better distinguish two types of dependencies: 1) an API requires two parameters returned by both two APIs ($fn_i\to par_j \& fn_{i+1}\to par_{j+1}$), and 2) an API requires one parameter returned by either of the two APIs ($fn_i\to par_j | fn_{i+1}\to par_j$). Figure~\ref{fig:ADGraph} demonstrates the API dependency graph for \texttt{toylib} discussed in Figure \ref{fig:example}. The graph contains five API nodes $FN_5=\{$\texttt{f1,f2,f3,f4,f5}$\}$ and two parameter nodes $PAR_2=\{$\texttt{s1,s2}$\}$. \texttt{s1} and \texttt{s2} are two non-primitive type parameters required to call \texttt{f4}; \texttt{s2} is one non-primitive type that can be produced by either \texttt{f2} or \texttt{f3}.

We further define two kinds of special API nodes: \textit{start node} that consumes no non-primitive parameters and \textit{end node} that returns no non-primitive values. In Figure~\ref{fig:ADGraph}, \texttt{f1}, \texttt{f2}, \texttt{f3} are start nodes and \texttt{f5} is an end node. \texttt{f4} is neither a start node nor an end node. Besides, an API node can be both a start node and an end node. 

\subsection{Validity of API Sequence} \label{sec:validity}
On an API dependency graph, an API node is \textit{reachable} if and only if it is a start node or all its required parameter nodes are reachable and the weights of consumer edges are satisfied. Similarly, a parameter node is reachable if at least one API node that can produce the parameter is reachable.

We say an API sequence $fn_0,...,fn_k$ is valid if $fn_0$ is a start node and each $fn_i (0< i\le k)$ is reachable given the subsequences of $fn_0,...,fn_{i-1}$. This is an essential requirement for synthesizing valid fuzz targets. For example, \texttt{\{f2,f5\}} is a valid sequence in Figure~\ref{fig:ADGraph}; \texttt{\{f1,f4\}} is invalid because \texttt{s2} is not reachable for reaching \texttt{f4}.

Note that our definition of a valid API sequence does not enforce a precedence relationship between each pair of adjacent API nodes. In this way, a sequence with duplicated subsequences (or duplicated APIs) or with multiple end nodes can also be valid. This is essential to deal with the move semantics in Rust. Briefly, move semantics mean the return value can be used only once. If an API has two parameters of the same type, we may call \texttt{f1} twice to serve the two parameters. Therefore, \texttt{\{f1,f1\}} is also a valid sequence.

\subsection{API Sequence Generation}
We generate valid API sequences by traversing the API dependency graph. According to the objective of fuzz target generation discussed in Section~\ref{sec:challenge}. The algorithm should meet three objectives: API coverage, efficiency, and effectiveness. To this end, we first employ BFS with pruning to obtain a set of API sequences that can cover the maximum number of APIs given a length threshold; next, we backward search valid sequences for uncovered APIs with lengths larger than the BFS threshold. Finally, we merge the two sets as one and further refine it to remove redundancies. 

\subsubsection{BFS with Pruning}
We choose BFS as a basic algorithm to generate valid sequences because fuzzing prefers short sequences. Other traversal algorithms like random walk are not suitable. 

Algorithm \ref{algorithm-bfs} introduces our basic algorithm. The main part of the algorithm is a while loop (line 6-17) that iterates over the length of the generated sequences and terminates once reaching a threshold. We start from a set with an empty sequence. In each iteration, we extend the sequences of length \texttt{i-1} with one more API node to generate new sequences (line 7-13). To ensure validity, we perform \texttt{ReachabilityTest} (line 11) for each candidate API as discussed in Section~\ref{sec:validity}. Besides, we employ an \texttt{EndNodeTest} function (line 8) and a \texttt{RedundancyTest} function (line 19) to optimize the sequence generation algorithm.  

\input{algorithm/bfs}

The \texttt{EndNodeTest} function intends to terminate one thread earlier once it reaches an end node. If we continuously append new APIs to a sequence with an end node, there must be another different sequence which is the same length and contains the same set of APIs and producer edges. For example, \texttt{\{f2,f5\}} is a sequence of \texttt{toylib}, and \texttt{f5} is an end node. If we append another API \texttt{f3} to the sequence, we will get \texttt{\{f2,f5,f3\}}, which contains the same set of APIs and producer edges with another equal-length sequence \texttt{\{f2,f3,f5\}}. For such cases, our algorithm will generate \texttt{\{f2,f3,f5\}} only and prune \texttt{\{f2,f5,f3\}} as redundant.

The \texttt{RedundancyTest} function is designed to filter another type of redundant sequences. Such sequences contain an API in the middle of the sequence that is not the dependency of any following API. For instance, \texttt{f1} is such an API in the sequence \texttt{\{f3,f1,f5\}}. The sequence is redundant since BFS has already generated another two shorter sequences \texttt{\{f1\}} and \texttt{\{f3,f5\}} that contain the same APIs and producer edges. For such cases, our algorithm will generate \texttt{\{f1\}} and \texttt{\{f3,f5\}} and prune \texttt{\{f3,f1,f5\}} as redundant (line 19 in Algorithm \ref{algorithm-bfs}).

\begin{figure*}[htbp]
    \centering
    \includegraphics[width = 0.9\linewidth]{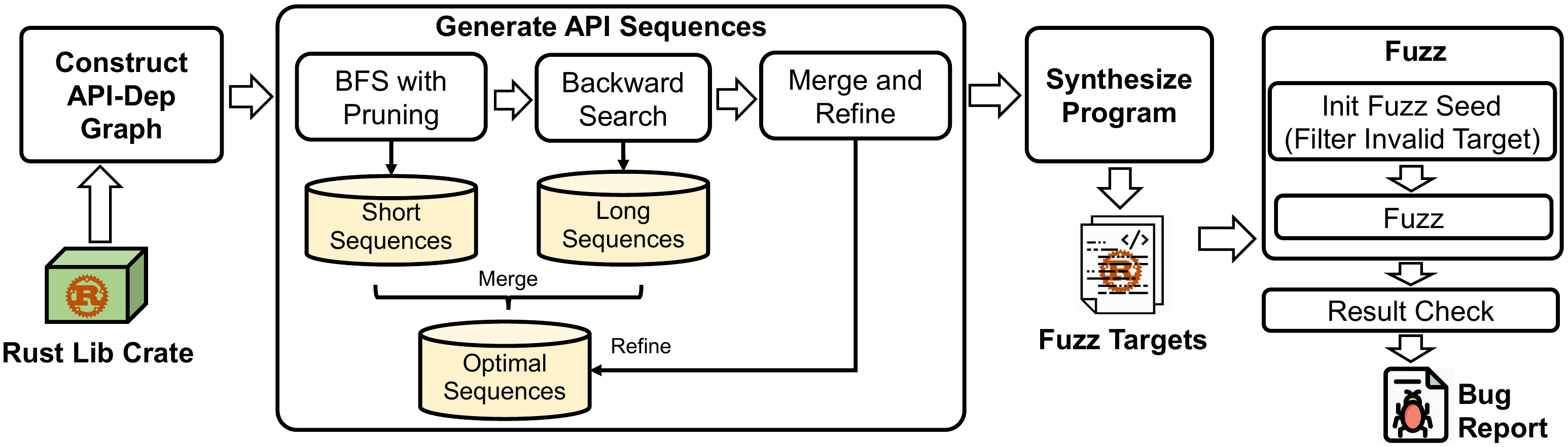}
    \caption{Overview of our workflow.}
    \label{fig:overview}
\end{figure*}

\subsubsection{Backward search} \label{subsubsection: backward search}

BFS with a threshold of sequence length cannot cover APIs that require longer sequences (deep-APIs). For example, an API that accepts several parameters may need a long call sequence to satisfy each parameter requirement. We observe that after running BFS with an arbitrary threshold three, only a few APIs cannot be covered. To reach these APIs, continuously searching for longer valid sequences with BFS would be inefficient. However, we can search the dependencies for these remaining uncovered APIs backward. 

For each uncovered API, our backward search approach finds its parameters directly based on already-generated sequences. For \texttt{toylib}, after performing BFS with a threshold one, we will generate three sequences \texttt{\{f1\}}, \texttt{\{f2\}}, \texttt{\{f3\}}. Then, we backward search for the dependencies of \texttt{f4}, which are \texttt{s1} and \texttt{s2}. To produce \texttt{s1} and \texttt{s2}, we can find a producer edge from \texttt{\{f1\}} to \texttt{s1} and another producer edge from \texttt{\{f2\}} to \texttt{s2}. Since all parameters of \texttt{f4} are satisfied, we can generate a new sequence \texttt{\{f1,f2,f4\}} covering \texttt{f4}. Similarly, we can generate a sequence \texttt{\{f2,f5\}} covering \texttt{f5}. We perform backward search iteratively until no more APIs can be covered.

\subsubsection{Merge and Refine}
Now, we merge the sequence sets generated by BFS and backward search and refine the set. The goal of our refinement is to select a minimum subset of sequences offering equivalent API coverage. This is in general a \textit{Set Covering Problem (SCP)}~\cite{SCP-complete}, which is NP-complete. 

To solve the problem, we employ a greedy algorithm~\cite{heuristic-algorithm}. Our algorithm selects a sequence that contributes most to the coverage until all APIs are covered. In particular, we prefer a candidate sequence A over B based on the following rules. 

\begin{itemize}
    \item A covers more new nodes than B.
    \item A and B cover the same number of new nodes, but A covers more new edges than B.
    \item A and B cover the same number of new nodes and edges, but A is shorter than B.
\end{itemize}

If there are multiple sequences with equal contribution based on the above rules, we randomly select one from them.

Suppose \texttt{\{f3\}} and \texttt{\{f3,f5\}} are two candidate sequences for selection. If \texttt{f5} or the production edge from \texttt{f3} to \texttt{s2} (the parameter node of \texttt{f5}) has not been covered, we select \texttt{\{f3,f5\}} because it contributes more to coverage. Otherwise, we select \texttt{\{f3\}} because it is shorter.

After applying all the sequence generation steps to \texttt{toylib}, we finally obtain two sequences as shown in the boxes in Figure \ref{fig:target1} and \ref{fig:target2} (or two similar sequences).

\section{Implementation}\label{section:Implementation}

We implement a prototype tool \name{} to automatically generate fuzz targets for Rust libraries. Our tool is based on Rust 1.46.0-dev and contains about 6K lines of Rust codes. 

Figure \ref{fig:overview} overviews the workflow of \name{} that inputs a Rust lib crate and outputs a set of fuzz targets. There are three essential components. The first is to construct an API dependency graph based on API signatures. The second is to generate valid sequences by traversing the graph. The third is to synthesize Rust programs ready to compile from valid sequences. 

Next, we discuss the implementation details of each component.

\subsection{Construct API Dependency Graph} \label{subsection:ADgraph}

We construct an API dependency graph in two stages. 

First, we extract all public API signatures from the source code. We base \name{} on an existing tool, rustdoc~\cite{Rustdoc}, which is an official doc tool to generate API documentation for Rust projects. Internally, rustdoc calls rust compiler to compile the crate and extract API information from the compilation results. We directly add our codes into librustdoc. 

After we extract all public API signatures, we infer dependencies among APIs based on type inference to build dependency graph. We define \textit{call type} for type inference. If there is a call type from type A to type B, A and B are the same type or A can be converted to B. All call types can be seen in Table \ref{tab:call-types}. Note that call types can be nested, \textit{e.g.,} there is a call type from \texttt{Option<T>} to \texttt{\&T}, which is \textit{Borrowed reference (Unwrap option)}.

\input{tables/call-type}

Call types defined in Table \ref{tab:call-types} can be divided into two categories. The first seven are general call types (from \textit{Direct call} to \textit{Dereference raw pointer}). The last three are specific call types for two types, \texttt{Result} and \texttt{Option} (from \textit{Unwrap result} to \textit{To option}). \texttt{Result} and \texttt{Option} are two heavily-used enumeration types in Rust. \texttt{Result} is used to return recoverable errors. \texttt{Option} indicates a value that may be empty. These two types are used to wrap a normal type, \textit{e.g.,} \texttt{u8} is an 8-bit integer type, while \texttt{Option<u8>} can be a normal \texttt{u8} or \texttt{None}. We can get the real \texttt{u8} type by unwrapping \texttt{Option<u8>} if it is not \texttt{None}. If these two types are not specially treated, it will greatly affect the validity of type inference. 
 
More wrapper types can be treated specially, \textit{e.g.,} \texttt{Box}. Special treatment of these types helps infer dependencies more accurately. \name{} can support this by adding a new call type.
 
\subsection{Generate API Sequences}
 
After we construct an API dependency graph, we generate valid sequences by traversing the graph. The main approach is mentioned in the previous section. Here are more details.

\begin{figure}[ht]
    \centering
    \includegraphics[width=\linewidth]{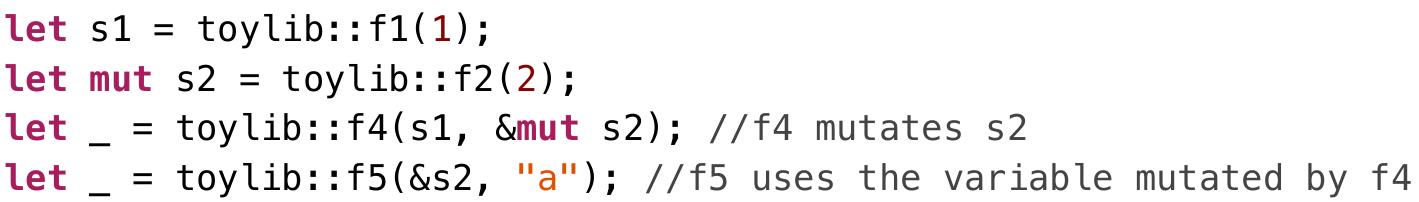}
    \caption{A mutation case on crate \texttt{toylib}.}
    \label{fig:mutate}
\end{figure}

API dependency graph can only capture all possible data flow dependencies, but this is not enough for generating all possible sequences. Figure \ref{fig:mutate} demonstrates such a case. Since the return value of \texttt{toylib::f4} is never used, this sequence will be removed by $RedundancyTest$ in previous section. However, \texttt{toylib::f5} uses the variable mutated by f4. So, \texttt{toylib::f4} is not a redundant API. 

To deal with this case, we integrate $EndNodeTest$ and $RedundancyTest$ with mutation analysis. If an API has no return value or returns a primitive type, it is an end node according to the definition in the previous section. However, if the API can mutate the value of a variable (like setter methods in Java), it is not appropriate that we treat this API as an end node. The $EndNodeTest$ is based on the fact that the API behind an end node can exchange orders with other end nodes. This fact does not hold if the end node can mutate the value. So we redefine an API that satisfies the following two requirements as an end node: 1) The API has no return value or returns a primitive type. 2) The API does not mutate the value of any variable that another API returns.

We modify $RedundancyTest$ similarly: if neither the return value of an API nor variables whose values are mutated by this API are used by other APIs, this API can be called a dead node. Sequences containing such a dead node are redundant and can be filtered. 

Determining whether an API can mutate a value in Rust programs is easy. Rust has a keyword \texttt{mut}. Only if the parameter of an API is modified with the keyword \texttt{mut}, the API has the ability to mutate a value (\textit{e.g.,} \texttt{toylib::f4} can mutate the value of \texttt{s2}). After modifying the definitions of $EndNodeTest$ and $RedundancyTest$, we can still use the algorithms in the previous section. Interior mutability in Rust, allowing unsafe code in a function to mutate the value of a parameter even if the parameter is not modified with \texttt{mut}, may affect the effectiveness of mutation analysis. However, interior mutability does not affect the validity of sequences generated by \name{}. So we ignore the limitation in the current implementation of \name{}.

\begin{figure}[ht]
    \centering
    \includegraphics[width=\linewidth]{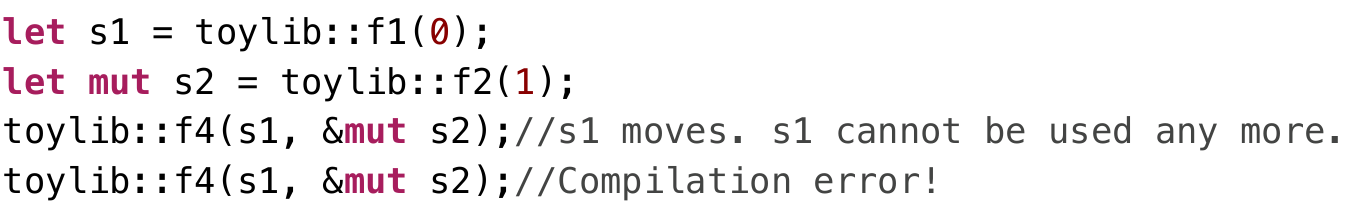}
    \caption{A simple move case on crate \texttt{toylib}.}
    \label{fig:move-case}
\end{figure}

To generate sequences that can pass the compiler's check, we take Rust’s language features into account. Ownership is Rust’s most unique feature. In Rust, each value has only one variable as its owner which has the exclusive ownership of the value. The ownership can be transferred among variables, which is called \textit{move}. If a variable transfers the ownership, the variable cannot visit the value anymore. Figure \ref{fig:move-case} demonstrates a simple move case. \texttt{s1} defined in line 1 is used by \texttt{toylib::f4} in line 3. \texttt{s1} transfers the ownership in line 3. We cannot use \texttt{s1} in line 4 any more otherwise it will report a compilation error.

Passing a value to a function may result in the value moved or copied, so we should distinguish between moving and copying cases. If a value is moved, we add a tag to the variable which owns the value and do not use this variable in the following statements. Whether a value moves depends on the combination of call type and variable type. Considering that there are many call types (call types can be nested) and variable types, it is not easy to distinguish moving and copying cases correctly. In practice, we adopt a conservative strategy. We select some combinations of variable types and call types that we think are commonly used in API calls and mark these combinations as copying cases if they do be. For example, if the variable type is immutable reference and the call type is \textit{Direct Call}, we will mark this case as a copying case. Except for several cases marked as copying cases, we assume that all other cases are moving cases. 

Reference is also carefully dealt with. Variables without ownership can visit a value by having a reference. There are two categories of reference, immutable reference (shared reference) and mutable reference (exclusive reference). To avoid data race, immutable reference and mutable reference of the same value cannot coexist. Multiple mutable references of the same value are not allowed to coexist too. We add a tag to variables that have a reference to make sure that the rules about references are correctly followed. Currently, we do not take special treatment to the lifetime bounds of references, which may result in generating invalid sequences. We will add check of lifetime in our following work.
 
\subsection{Synthesize Program} \label{subsection: Generating fuzz targets}

We synthesize Rust programs ready to compile from valid sequences. We implement several stages in \name{} to pass the check of the compiler, \textit{e.g.,} adding mut tag, adding import statements and so on. In this section, we focus on our method to deal with primitive types.
 
We get all primitive parameters from the input buffer of fuzz targets. We split this buffer into slices of different sizes. Then we convert each slice to a primitive value. This conversion relies on a series of template functions we write in advance. We currently do not support some primitive types, \textit{e.g.,} string with static lifetime or array with a static length that cannot be provided by the fuzzer engine.
 
To decide how to split the input buffer, we divide all primitive types into two categories. One is with a fixed length, such as u8 and char. The other is with a dynamic length, such as string and slice. We first assign a slice from the buffer to each fixed-length variable. The remaining part of the buffer will be divided equally among all dynamic-length variables. 

When we merge and refine the sequences after backward search, we skip sequences without primitive parameters as input since fuzzing these sequences is useless. We also give preference to sequences with primitive parameters of dynamic length. These sequences have larger input space to search and intuitively are more likely to contain errors.

\section{Evaluation Experiment}\label{section:experiment}
This section evaluates \name{} by studying three research questions.

\begin{itemize}
    \item \textbf{RQ1:} Can \name{} meet the four objectives (\textit{i.e.,} validity, API coverage, efficiency, and effectiveness) of fuzz target generation?
    \item \textbf{RQ2:} How does each step of our sequence generation algorithm help generate high-quality fuzz targets?
    \item \textbf{RQ3:} Can \name{} outperform the state-of-the-art approach?
\end{itemize}

\subsection{Experiment Setting}
We select 14 popular and high-quality Rust crates for fuzzing experiments, including eleven from crates.io and three from GitHub. All these crates have been downloaded over ten million times or received more than 4000 stars. Meanwhile, all these crates have well-written unit tests and stable releases. All these crates manage their dependencies by \texttt{cargo}, \textit{i.e.,} running \texttt{cargo build} for these crates will return success without manually downloading dependencies. We avoid selecting certain libraries due to our current limitations (see Section \ref{section:limitation}), \textit{e.g.,} libraries containing many asynchronous APIs. We download their latest version for our experiments. Table~\ref{tab:results} lists the names of these crates. Their API numbers vary a lot, ranging from 12 to 212. 

When generating fuzz targets with BFS, we choose three as the threshold of sequence length. The threshold cannot be large due to path explosion. For instance, the running time to generate fuzz targets on a crate \texttt{time} will exceed one hour if we set the threshold to four. Meanwhile, the threshold cannot be too small because we cannot explore different API combinations in a very short sequence. Three is a best practice according to the code samples in Trophy-Case to reproduce bugs. We find that most bugs can be reproduced by calling library APIs within three times.

Before fuzzing each fuzz target, we generate 500 random inputs and execute the target with each input. If all inputs lead to program crashes, the fuzz target is considered as an invalid target and removed. The rest inputs will be employed as seeds for further fuzzing.

We use \texttt{afl.rs} from rust-fuzz~\cite{RustFuzz} to compile and fuzz these targets. \texttt{afl.rs} is a Rust binding for AFL++~\cite{fioraldi2020afl++}, which is AFL with community patches. We set the environmental variable \verb|AFL_EXIT_WHEN_DONE| = 1, so the fuzzing process will exit automatically when all branches are visited. We set \verb|AFL_NO_AFFINITY| = 1, so we can fuzz more targets parallel than the number of our cores. All other environmental variables are set with default values. We fuzz each crate for 24 hours unless AFL itself terminates. After that, we obtain a set of crashes and refine the result with afl-tmin and afl-cmin. In particular, these tools can merge crash reports with equivalent edge coverage. 

We manually check each crash report to determine whether it is a real API bug. To this end, we reproduce each crash to analyze why the crash happens. If a crash is caused by an explicit panic function call or is documented as a known panic case, we do not treat it as a bug. Otherwise, the crash should indicate a bug. 

All our experiments are conducted on a Dell T640 server with 2 Intel Xeon Gold 6242R 3.10GHz 20-core CPUs and 256GB memory.

\input{tables/result}

\subsection{\textbf{RQ1}: Can \name{} meet the four objectives of fuzz target generation?}

We discuss the performance of \name{} with respect to the objectives of validity, API coverage, efficiency, and effectiveness. 

\subsubsection{Validity and API Coverage}

\begin{figure}[ht]
    \centering
    \includegraphics[width=\linewidth]{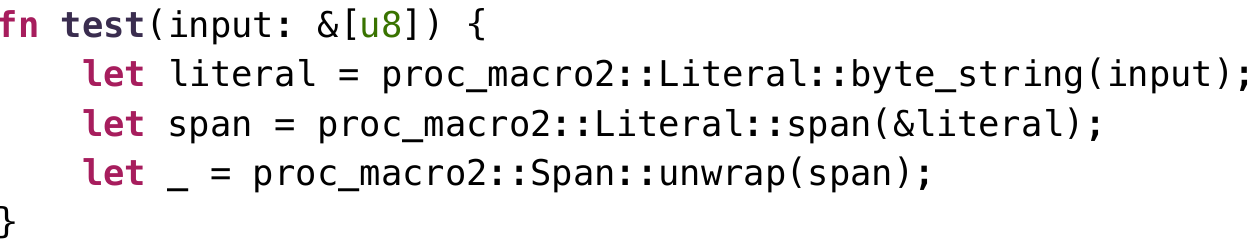}
    \caption{An invalid target on \texttt{proc\_macro2}.}
    \label{fig:invalid}
\end{figure}

Table~\ref{tab:results} overviews our experimental results. We generate 7 to 118 targets for each library, and the API coverage ranges from 0.27 to 0.92. Since we generate sequences by traversing the API dependency graph, all parameters of each API are settled correctly, and all sequences are syntax-correct and can be compiled successfully. Among all fuzz targets, there are only 4/35 invalid targets for the crate \texttt{clap} and 1/33 for \texttt{proc-macro2}. The main reason is that our synthesized program does not follow the contextual requirement for using particular APIs. Figure \ref{fig:invalid} demonstrates the issue with an invalid target for \texttt{proc\_macro2}. The API \texttt{proc\_macro2::Span::unwrap} is used to convert \texttt{proc\_macro2::Span} to \texttt{proc\_macro::Span}. However, since \texttt{proc\_macro::Span} only exists within the context of procedural macro (Rust's syntax extensions), \texttt{proc\_macro2::Span::unwrap} will always panic if we call it outside procedural macros.

\begin{figure}[ht]
    \centering
    \includegraphics[width=\linewidth]{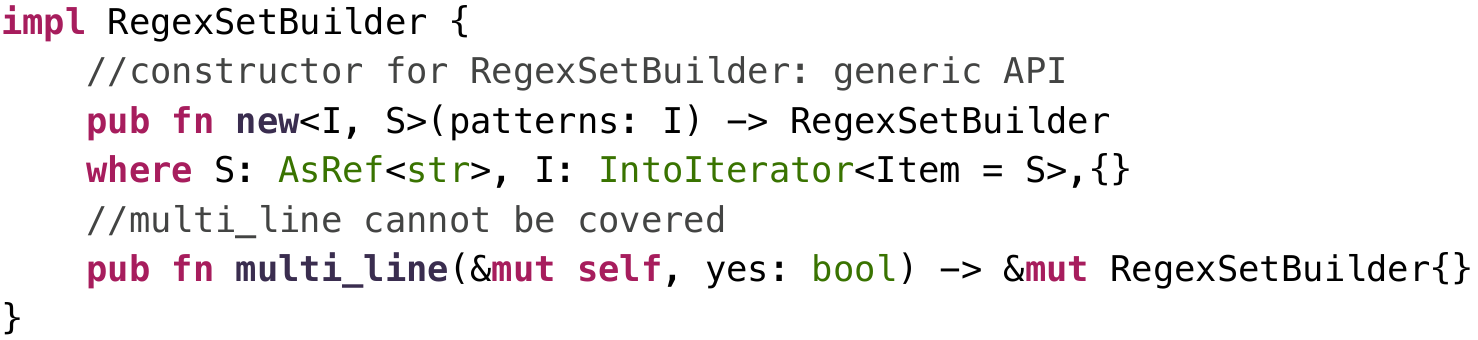}
    \caption{An uncovered API in \texttt{regex}.}
    \label{fig:notcover}
\end{figure}

\name{} achieves API coverage ranging from 0.27 to 0.92. The remaining APIs cannot be covered mainly because their parameters cannot be served by other APIs of the same crate. A large portion of these APIs depends on generic APIs that involve dynamic types. Figure \ref{fig:notcover} demonstrates such an uncovered API \texttt{multi\_line}. The API requires a parameter of type \texttt{RegexSetBuilder} (represented by \texttt{self}). However, the only constructor method for \texttt{RegexSetBuilder} is a generic API. We currently do not support synthesizing valid programs with generic parameters. Section \ref{section:limitation} discusses more on this issue.

\subsubsection{Effectiveness}

Our results show \name{} is effective in finding bugs in libraries. We find 30 previously-unknown bugs in total in seven libraries. All bugs have been included in Trophy-Case. We manually check each crash and collect bugs with different API call paths. We submit issues about all bugs to the GitHub repository of these libraries. 24 bugs are confirmed by the repository maintainers before we submit our paper. The types of bugs are shown in Table \ref{tab:bugs}. There are five kinds of bugs in Table \ref{tab:bugs}, which are all common bugs in Rust. 

\input{tables/bugs}

\begin{itemize}
    \item Range: out-of-bound error
    \item Unwrap: call \texttt{unwrap} on \texttt{Err()} or \texttt{None}
    \item Arith: arithmetic overflow
    \item UTF-8: problems with UTF-8 strings
    \item Unreachable: visit codes with macro \texttt{unreachable!()}
\end{itemize}

False positive is a key concern when we fuzz many targets. False positive crashes are inevitably introduced to the results because the library developers may determine to panic the program under certain circumstances, \textit{e.g.,} when the library API receives invalid parameters. In practice, the false positive crashes can be determined at a glance due to clear output information. Our preference to short sequences also facilitates checking and analyzing each crash report. In our experiment, all crashes on one fuzz target do point to the same bug (or the same false positive reason). To check all crashes of one crate takes a graduate student ranging from one to five hours.

\begin{figure}[ht]
    \centering
    \includegraphics[width=\linewidth]{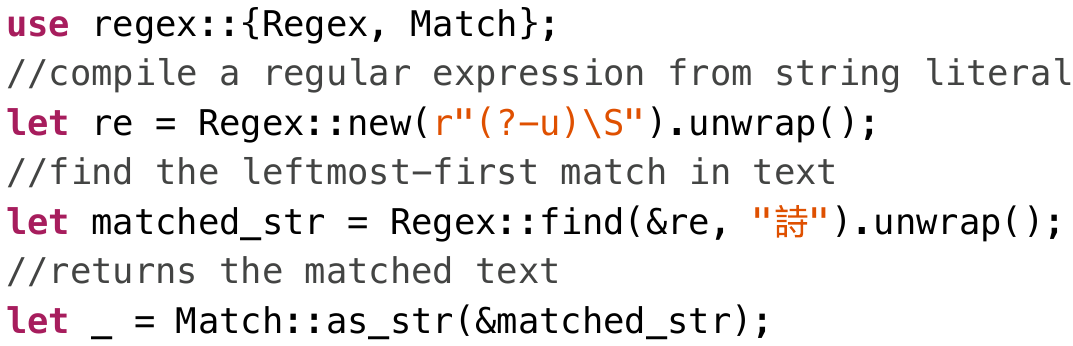}
    \caption{Code to reproduce one bug we found in crate \texttt{regex}.}
    \label{fig:reproduce-regex}
\end{figure}

We present one library example as a case study. \texttt{Regex} is a crate from Rust's official team. It is one of the most important crates for Rust and is imported by more than three thousand projects on crates.io. \texttt{Regex} has well-written unit tests and is continuously fuzzed with OSS-fuzz~\cite{oss-fuzz}. \name{} finds ten new bugs in regex. Six bugs are related to improperly dealing with unexpected inputs. The remaining four bugs are logical errors. The code to reproduce one bug is in Figure \ref{fig:reproduce-regex}.

\begin{figure*}[htb]
    \centering
    \includegraphics[width=\linewidth]{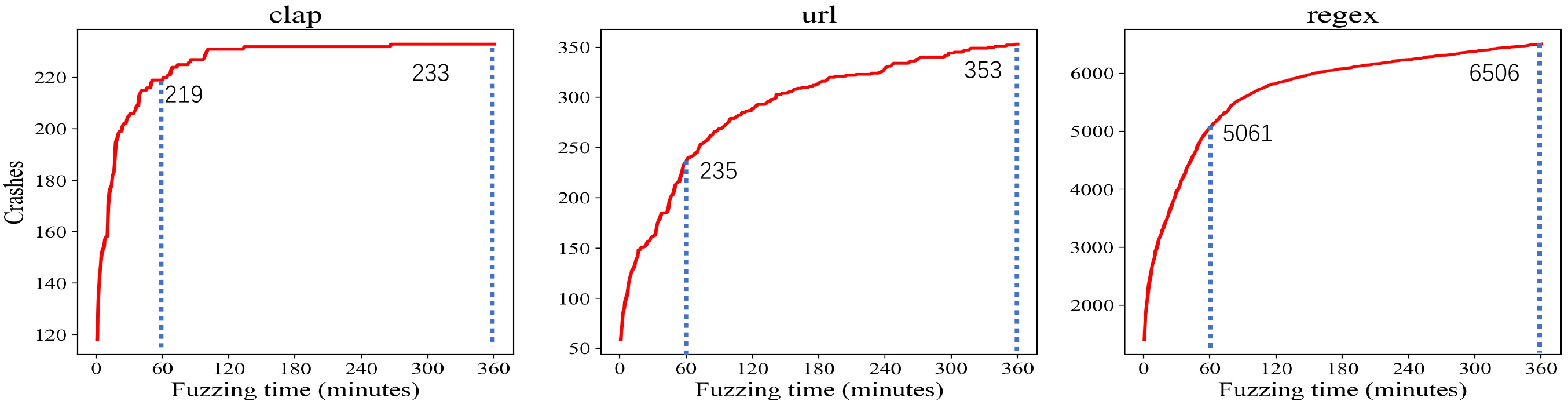}
    \caption{The number of crashes of fuzzing three crates where we find bugs in first six hours. We sample every minute.}
    \label{fig:effiency}
\end{figure*}

\input{tables/comparison-search-strategy}
\input{tables/algorithm-efficiency}
\input{tables/comparison}

The reason behind this bug is that \texttt{Regex} is not allowed to match invalid UTF-8 text. \texttt{Match::as\_str} always assumes it accepts a valid UTF-8 text, otherwise it will panic. However, \verb|\S| can match any invalid UTF-8. \texttt{Regex::find} matches an invalid UTF-8 text in our case. So \texttt{Match::as\_str} panics and reports an invalid UTF-8 error. The bug roots in \texttt{Regex::new}, where the given regex should not be allowed to compile.

Note that although the bug roots in \texttt{Regex::new}, we detect panic behaviors in \texttt{Match::as\_str}. If we do not construct such an API sequence and only fuzz \texttt{Regex::new}, we cannot detect the bug due to no abnormal behaviors are observed. We need to combine the APIs in a specific way. Besides, the bug can only be triggered by the input string with \verb|\S|. Fuzzing can efficiently find the exact input. By combing fuzzing and API sequence generation, \name{} is effective in finding these bugs.

\subsubsection{Efficiency}

We reduce the size of the fuzz target set by pruning and refining. The average ratio of APIs to fuzz targets is 1.43. Considering the length of most sequences is smaller than or equal to three, the set of fuzz targets is compact. Table \ref{tab:comparison-search-strategy} shows in most cases, \name{} visits each API for no more than two times on average. This meets our design objective.

However, we still generate up to 118 fuzz targets for each library. Previous work like Fudge~\cite{babic2019fudge} presents only a few targets for each library. So it is necessary to evaluate the efficiency of fuzzing by experiment.

The fuzzing efficiency varies greatly among different libraries. As shown in Table \ref{tab:results}, the average run time ranges from 0.27 to 24 hours. For libraries like \texttt{time} and \texttt{xi-core-lib}, all fuzzing processes exit in several hours (about 4 and 3 hours respectively). For libraries with complex APIs, \textit{e.g.,} \texttt{url}, no fuzzing processes exit in 24 hours. The fuzzing efficiency may be more related to the complexity of library APIs rather than the number of fuzz targets.

Figure \ref{fig:effiency} shows the relation between the number of reported crashes and time. Newly-discovered crashes in each minute are fewer and fewer. The reported crashes in the first hour take 0.67 to 0.94 of those in the first six hours. For most libraries, newly-reported crashes after six hours are few. This indicates \name{} is efficient for fuzzing as the number of crashes converges in a not very long time.

\subsection{\textbf{RQ2:} How does each step of our sequence generation algorithm help generate high-quality fuzz targets?}

The basic goal of our algorithm is to generate a set of valid sequences to cover all APIs. Three optimization goals are (1) to minimize the set size (2) to maximize producer edge coverage (3) to avoid repeatedly visiting the same API.

Combining BFS with backward search enables us to cover all APIs efficiently. It takes less than 30 seconds to generate targets for each library if the library is compiled.

Our pruning and refining algorithms reduce the size of the sequence set. As shown in Table \ref{tab:algorithm-effiency}, the number of sequences greatly decreases after we integrate pruning and refining algorithms into BFS. The ratio of sequences generated by BFS to fuzz targets after refining ranges from 28 to 3241, \textit{i.e.,} we achieve equivalent API coverage with a much smaller set. We can avoid fuzzing the same API too many times.

Backward search is efficient for covering deep-APIs. The longest sequence we generate by backward search is of length seven. It is hard for only BFS to generate this long sequence due to path explosion. On our server, we cannot generate this long sequence with only BFS before running out of memory. Besides, we cannot determine which API cannot be covered with only BFS. Backward search can cover 26 more APIs in 5 out of 14 libraries after BFS. This number (26) is much smaller comparing to total API numbers (1435), which confirms our previous observation that only a few APIs cannot be covered by BFS. Among all 30 bugs we found, one bug is reproduced by calling four library APIs. We cannot find this bug without backward search. 

BFS is useful to combine APIs in different ways. A possible algorithm to cover all APIs is to only leverage backward search to generate sequences from an empty sequence. For each unvisited API, backward search tries to find dependency for each parameter of the API from already-generated sequences. Table \ref{tab:comparison-search-strategy} compares backward search, backward search with refining and \name{} on the quality of generated targets based on our three optimizing goals. To cover all APIs, backward search requires generating more targets and visiting the same API repeatedly for more times. Besides, backward search can cover fewer producer edges. Even after performing our refining method on the results of backward search to remove redundant targets, \name{} performs better. We limit our use of backward search to only a few deep-APIs to ensure the quality of generated targets. The reason is that BFS generates all possible short sequences. When we refine the result, we have more sequences to select from. So we can obtain targets with higher quality. 

\subsection{\textbf{RQ3:} Can \name{} outperform the state-of-the-art approach?}

There is no available fuzz target generator on Rust now. The closest one is Fudge~\cite{babic2019fudge}. We select Fudge as our first baseline. Fudge can generate fuzz targets for C/C++ libraries automatically. Fudge has earned great success and led to hundreds of bug fixes. Fudge extracts code snippets as targets by finding the minimum data flow and control flow dependencies of library APIs in client codes. Fudge extracts code snippets from the whole Google code bases. Since we cannot visit Google code bases, we adopt Fudge's algorithm on two available code bases: \textit{(i).} ten client libraries with most downloads on crates.io using APIs from the libraries to fuzz (FLC) and \textit{(ii).} the unit tests from the library to fuzz (FLUT). We extract codes manually with Fudge's algorithm and drop duplicate targets with the same API call sequences.

We also implement random walk as our second baseline. One possible alternative to generating targets is random walk. Random walk starts from an empty sequence. For each step, we pick up a random sequence from previously-generated sequences and a random API from all library APIs. If the API can be added to the sequence to generate a valid sequence, we generate a new sequence. We perform random walk until we generate the same number of targets as generated by \name{}.

For \name{}, we fuzz each crate by a threshold of 6 hours. For other methods, we ensure the total time threshold to fuzz a library is the same as \name{}. We compare \name{} with other methods on the effectiveness of bug finding. The results can be seen in Table \ref{tab:comparison}.

The comparison results can be seen in Table \ref{tab:comparison}. Among four methods, random walk performs most poorly. Randomly-generated targets usually cover the same APIs and paths, so fuzzing these targets leads to the fewest bug findings. FLC behaves better than random walk but still poorly. We find that client codes usually limit their usage of library APIs to some most common ones, so we can only extract very few targets. Client codes tend to use APIs with complex logic which are more error-prone to some extent. So FLC can find more bugs than random walk. FLUT achieves higher API coverage and finds more bugs than the first two methods. Library unit tests combine library APIs in different ways to ensure APIs behave properly. The codes in unit tests cover most error-prone cases. FLUT can find some bugs beyond the capacity of the current version of \name{}, \textit{e.g.,} bugs involving third-party APIs and advanced Rust features like closure. However, similar to client codes, unit tests emphasize more on the combination of commonly-used APIs. So, FLUT fuzzes commonly-used APIs repeatedly while missing bugs in other rarely-used APIs. To improve the performance of Fudge-like methods, we require better code bases to capture the usage of library APIs. \name{} behaves best for we directly generate targets to achieve high API coverage from library source codes. We will not miss any API that may go wrong.

\subsection{Limitations and Future Work} \label{section:limitation}

\name{} still has some limitations, and we leave these issues as our future work.

Firstly, the API coverage of \name{} can be further improved by supporting polymorphic APIs. These APIs involve polymorphic types, such as generics and trait object. Since we infer dependencies only according to concrete types, we cannot infer dependencies for polymorphic APIs now. We will consider supporting typical usage patterns of polymorphic APIs in the future, such as trait object or generics with trait bound by exhausting all possible types defined in the crate.

Secondly, the current version of \name{} only considers functions or methods as library APIs for fuzzing. It cannot support macro APIs. We will consider supporting macro in the future. \name{} currently also does not support asynchronous APIs. The execution of asynchronous APIs depends on a specific runtime, such as async-std or Tokio~\cite{async-rust}. We will investigate how asynchronous APIs influence the effectiveness of fuzzing in our following work.

Finally, \name{} only supports general fuzzing tools like AFL and honggfuzz. We cannot support in-process fuzzers like libFuzzer because we introduce \texttt{std::process::exit} in our template functions to simplify current implementation of \name{}. We will consider replacing the use of \texttt{std::process::exit} to support in-process fuzzers. Besides, general fuzzing tools are inefficient to deal with highly structured inputs~\cite{fioraldi2020weizz}~\cite{mathis2020learning}. We will investigate how \name{} performs when we fuzz programs with different fuzzers.

\section{Related work} \label{section:related work}

Fuzzing tools such as AFL~\cite{AFL}, libFuzzer~\cite{LibFuzzer}, honggfuzz~\cite{Honggfuzz} have attracted much attention in recent years. Existing investigations mainly aim to improve the efficiency and effectiveness of such tools, such as AFLFast~\cite{AFLFast2016}, Driller~\cite{stephens2016driller}, and Fairfuzz~\cite{lemieux2018fairfuzz}. All these tools require well-defined fuzz targets. 

There exists some work about fuzz target generation. Fudge~\cite{babic2019fudge} extracts short sequences from client projects as fuzz targets. The mechanism highly relies on high-quality code bases to generate fuzz targets. We have shown the limitations of Fudge with several crates in our comparison study. FuzzGen~\cite{fuzzgen2020} builds $A^{2}DG$ from existing code bases to represent the data flow and control flow dependencies and generate fuzz targets based on the $A^{2}DG$. FuzzGen also relies on high-quality code bases. \cite{kelly2019case} introduces a case study about automatically-generated fuzz targets intended for C programs. Their method to get primitive parameters from the input buffer is similar to ours. However, ~\cite{kelly2019case} only fuzzes a single API rather than combining APIs in different ways. RESTler~\cite{Restler} is another work that generates fuzz targets for RESTful APIs, and it is based on BFS. However, RESTful applications do not have strong type requirements as Rust.

There are several automated unit test generation or program synthesis approaches related to fuzz target generation. For example, Randoop~\cite{randoop} automatically generates valid method call sequences incrementally by using feedback information. RecGen~\cite{RecGen} uses mutation analysis to better select sequences and increase code coverage.  Concolic execution is another way to improve the performance of unit test generation~\cite{garg2013feedback}. SyRust~\cite{syrust21} is a testing framework for Rust libraries. SyRust synthesizes large numbers of programs for a small set of library APIs and does not mutate the input value of each program. All these approaches are not directly applicable to the fuzz target generation problem for Rust libraries.

\section{Conclusion}\label{section:conclusion}

To conclude, this work proposes an automated fuzz target generation approach for Rust library fuzzing based on API dependency graph traversal. Our approach employs BFS with pruning and backward search to generate valid API sequences and refines the sequences to obtain an optimal sequence set. We have implemented a prototype \name{} and show that it can successfully address the requirements of fuzz target generation for Rust libraries, including validity, API coverage, effectiveness, and efficiency. Considering the characteristic of Rust and its reliability requirement on library APIs, such tools are urgently needed and would play an essential role in facilitating the development of the community. 

\section*{Acknowledgment}

This work was supported by the National Natural Science Foundation of China (Project No. 61971145). Hui Xu is the corresponding author.

\bibliographystyle{IEEEtran}
\bibliography{IEEE-reference}

\end{document}

%% file: algorithm/bfs.tex
\begin{algorithm}
    \caption{BFS with pruning.}\label{algorithm-bfs}
    \SetAlgoNoLine
    \SetKwInOut{Input}{Input}
    \SetKwInOut{Output}{Output}
    \Input{A graph $G(FN,PAR,PE,CE)$,\\ 
    Sequence length threshold $max\_len$}
    \Output{A set of sequences $S_{BFS}$}
    
    \BlankLine
    \texttt{Init:} \\
    \quad   $S_{BFS},S_{new}\leftarrow \emptyset$ \;
    \quad   $S_{last} \leftarrow \{empty\_sequence\}$ \;
    \quad   $i$ $\leftarrow$ 1\;
     \BlankLine
    \texttt{BFS with Pruning:} \\
    \quad \While{$i$ $\le$ $max\_len$} {
    \quad    \For{$seq$ in $S_{last}$}{
     \quad     \If{EndNodeTest($seq$)}{
     \quad        continue\; }
    \quad      \For{$fn$ in $G\to FN$}{
    \quad         \If{ReachabilityTest($G$, $seq$, $fn$)}{
    \quad            $new\_seq$ $\leftarrow$ $seq$.append($fn$)\;
    \quad            $S_{new}$.add($new\_seq$)\;
               }}}
      \quad   $S_{BFS}$.append($S_{new}$)\;
      \quad   $S_{last} \leftarrow S_{new}$\;
     \quad    $S_{new} \leftarrow \emptyset$\;
     \quad    $i$ $\leftarrow$ $i$ + 1 \;
    }
    
    \quad  \For{$seq$ in $S_{BFS}$}{
    \quad       \If{RedundancyTest($seq$)}{
    \quad           $S_{BFS}$.remove($seq$);
            }}
\end{algorithm}

%% file: tables/call-type.tex
\begin{table}[ht]
  \centering
  \caption{Call types for type inference. T, E are any Rust type.}
  \label{tab:call-types}
  \begin{tabular}{lll}
    \hline
    Call type&Example\\
    \hline
    Direct call & T$\to$T \\
    Borrowed reference & T$\to$\&T \\
    Mutable borrowed reference & T$\to$\&mut T \\
    Const raw pointer & T$\to$ $\ast$const T  \\
    Mutable raw pointer & T$\to$ $\ast$mut T \\
    Dereference  borrowed reference & \&T$\to$T \\
    Dereference raw pointer & $\ast$const T$\to$T \\
    Unwrap result & Result$\langle$T ,E$\rangle$ $\to$T \\
    Unwrap option & Option$\langle$T$\rangle$ $\to$T \\
    To option & T$\to$Option(T) \\
  \hline
\end{tabular}
\end{table}

%% file: tables/result.tex
\begin{table*}[htbp]
    \centering
    \caption{Experiment results on 14 libraries. The top eleven are from crates.io. The last three are from Github. }
    \begin{tabular}{c|c|c|c|c|c|c|c|c}
    \hline
    Crate name & version & API numbers & API coverage & \makecell[c]{Fuzz targets \\ (invalid)} & Finished targets & \makecell[c]{Average run time \\ (hours) } & \makecell[c]{Crashes \\ (after cmin)} & Bugs\\
    \hline
    clap & 2.33.3 & 85 & 0.78 & 35(\textcolor{red}{4}) & 27 & 3.24 & 236(28) & 2 \\
    proc-macro2 & 1.0.24 & 60 & 0.65 & 33(\textcolor{red}{1}) & 16 & 20.42 & 451(103) & 0\\
    serde\_json & 1.0.61 & 89 & 0.46 & 25 & 24 & 1.20 & 0(0) & 0 \\
    http & 0.2.3 & 74 & 0.35 & 19 & 16 & 4.42 & 0(0) & 0 \\
    url & 2.2.0 & 74 & 0.91 & 61 & 0 & 24 & 422(75) & 2\\
    regex-syntax & 0.6.22 & 177 & 0.58 & 74 & 67 & 2.30 & 0(0) & 0\\
    regex & 1.4.3 & 129 & 0.74 & 67 & 20 & 16.84 & 7274(257) & 10\\
    semver & 0.11.0 & 12 & 0.92 & 7 & 0 & 24 & 147(1) & 1\\
    semver-parser & 0.10.2 & 15 & 0.80 & 10 & 7 & 13.56 & 0(0) & 0\\
    flate2 & 1.0.19 & 31 & 0.68 & 11 & 11 & 0.35 & 2(1) & 0\\
    time & 0.2.24 & 198 & 0.88 & 118 & 118 & 0.27 & 321(122) & 7\\
    bat & 0.17.1 & 91 & 0.27 & 20 & 2 & 21.76 & 0(0) & 0\\
    xi-core-lib & 0.4.0 & 212 & 0.27 & 34 & 34 & 0.53 & 34(10) & 4 \\
    tui & 0.13.0 & 188 & 0.37 & 48 & 48 & 3.12 & 105(39) & 4\\ 
    \hline
    \end{tabular}
    \label{tab:results}
\end{table*}


%% file: tables/bugs.tex
\begin{table}
  \centering
  \caption{The types of bugs we found by \name{}.}
  \label{tab:bugs}
  \begin{tabular}{c|c|c|c|c|c}
    \hline
    Crate & Range & Unwrap & Arith & UTF-8 & Unreachable\\
    \hline
    clap & 0 & 0 & 0 & 2 & 0\\
    url & 2 & 0 & 0 & 0 & 0\\
    regex & 6 & 0 & 2 & 2 & 0\\
    tui & 0 & 0 & 4 & 0 & 0\\
    time & 0 & 0 & 5 & 0 & 2\\
    xi-core-lib & 2 & 0 & 2 & 0 & 0\\
    semver & 0 & 1 & 0 & 0 & 0\\
  \hline
\end{tabular}
\end{table}

%% file: tables/comparison-search-strategy.tex
\begin{table*}[ht]
  \centering
  \caption{Comparison study of different search strategies for the quality of fuzz targets on five crates. }
  \label{tab:comparison-search-strategy}
  \begin{tabular}{c|ccc|ccc|ccc}
    \hline
    \multirow{2}*{Crate} & \multicolumn{3}{c|}{Backward search} & \multicolumn{3}{c|}{Backward search \&  Refine} & \multicolumn{3}{c}{\name{}}\\
    \cline{2-10}
    ~ & Targets & \makecell[c]{Covered edges} & \makecell[c]{ Average \# to\\ visit one API} & 
    Targets & \makecell[c]{Covered edges} & \makecell[c]{ Average \# to\\ visit one API} & 
    Targets & \makecell[c]{Covered edges} & \makecell[c]{ Average \# to\\ visit one API} \\
    \hline
    regex  & 94 & 94 & 2.30 & 81 & 94 & 2.08 & 67 & 94 & 1.92 \\
    clap  & 66 & 64 & 2.02 & 62 & 64 & 1.94 & 35 & 67 & 1.59 \\
    serde\_json  & 40 & 34 & 1.85 & 36 & 34 & 1.75 & 25 & 35 & 1.44 \\
    tui & 64 & 81 & 2.69 & 55 & 77 & 2.42 & 48 & 91 & 2.14 \\
    http & 26 & 19 & 1.73 & 22 & 19 & 1.58 & 19 & 21 & 1.54 \\
  \hline
\end{tabular}
\end{table*}

%% file: tables/algorithm-efficiency.tex
\begin{table}
  \centering
  \caption{The number of sequences on six crates after pruning and refining.}
  \label{tab:algorithm-effiency}
  \begin{tabular}{c|c|c|c|c}
    \hline
    Crate & BFS & \makecell[c]{EndNode- \\ Test} & \makecell[c]{Redundancy- \\ Test} & Refine\\
    \hline
    url & 1698 & 1520 & 171 & 61 \\
    regex & 2060 & 1909 & 352 & 67 \\
    time & 382401 & 285669 & 3960 & 118\\
    regex-syntax & 25748 & 18266 & 424 & 74\\
    serde\_json & 2178 & 2020 & 174 & 25\\
    clap & 5036 & 4758 & 3572 & 35\\
  \hline
\end{tabular}
\end{table}

%% file: tables/comparison.tex
\begin{table*}[htbp]
  \centering
  \caption{Comparison with baselines on three crates where \name{} found bugs about the effectiveness of bug finding. }
  \label{tab:comparison}
  \begin{tabular}{c|ccc|ccc|ccc|ccc}
    \hline
    \multirow{2}*{Crate} & \multicolumn{3}{c|}{Random walk} & \multicolumn{3}{c|}{FLC} & \multicolumn{3}{c|}{FLUT} & \multicolumn{3}{c}{\name{}}\\
    \cline{2-13}
    ~ & Targets & \makecell[c]{API \\ coverage} & Bugs & \makecell[c]{Targets} & \makecell[c]{API \\ coverage} & Bugs & \makecell[c]{Targets} & \makecell[c]{API \\ coverage} & Bugs & \makecell[c]{Targets} & \makecell[c]{API \\ coverage} & Bugs \\
    \hline
    url & 61 & 0.34 & 1 & 18 & 0.32 & 1 & 29 & 0.54 & 1 & 61 & 0.91 & 2 \\
    regex & 67 & 0.28 & 1 & 13 & 0.16 & 3 & 133 & 0.44 & 6 & 67 & 0.74 & 10 \\
    time & 118 & 0.38 & 0 & 15 & 0.11 & 0 & 124 & 0.43 & 5 & 118 & 0.88 & 7 \\
  \hline
\end{tabular}
\end{table*}